\documentclass{PoS}

\newcommand{\be}{\begin{eqnarray}}
\newcommand{\ee}{\end{eqnarray}}
\newcommand{\nl}{\nonumber\\}

\newcommand{\1}{\mathbb{I}}
\newcommand{\tr}[1]{\left\langle #1 \right\rangle}

\newcommand{\cL}{{\cal L}}

\newcommand{\cV}{{\cal V}}

\newcommand{\scpt}{S$\chi$PT}

\title{Chiral random matrix theory for staggered fermions}

\ShortTitle{Chiral random matrix theory for staggered fermions}

\author{\speaker{James C. Osborn}\\%
        Argonne Leadership Computing Facility,
        Argonne National Laboratory, Argonne, IL 60439, USA\\
        E-mail: \email{osborn@alcf.anl.gov}}

\abstract{
  We present a completed random matrix theory for staggered
  fermions which incorporates all taste symmetry breaking terms at
  their leading order from the staggered chiral Lagrangian.
  This is an extension of previous work which only included some of
  the taste breaking terms.
  We will also discuss the effects of taste symmetry breaking
  on the eigenvalues in the weak and strong taste breaking limits, and
  compare with some results from lattice simulations.
}

\FullConference{The XXIX International Symposium on Lattice Field Theory\\
  July 10-16, 2011\\
  Squaw Valley, Lake Tahoe, California}

\begin{document}

\section{Introduction}

Staggered fermions are an inexpensive and popular way of
discretizing quarks on a space-time lattice.  They preserve a chiral
symmetry with a local action at the expense of containing extra modes
that quadruple the number of quark species being simulated.  The
quadrupled species, called {\it tastes}, are mixed at nonzero lattice
spacing, but are expected to produce four independent quark flavors in
the continuum limit.  Evidence of this behavior can be seen in the low
eigenvalues of the staggered Dirac matrix.  At small lattice spacings
the eigenvalues tend to cluster into distinct quartets which represent
the four tastes.  As the lattice spacing is decreased, the eigenvalues
within a quartet will become more degenerate \cite{latev}.

In \cite{srmt} a Random Matrix Theory (RMT) for staggered
lattice fermions was introduced to describe the low eigenvalues of the
staggered Dirac operator.  The staggered RMT (SRMT) adds additional
terms to the standard chiral Random Matrix Theory that have the
appropriate symmetries of the lattice operator.  These additional
terms in the SRMT reproduce the known $O(a^2)$ (ignoring any extra
factors of $\alpha_s$ which we drop for convenience) taste breaking
terms that appear in the staggered chiral Lagrangian.

The SRMT is equivalent to the staggered chiral Lagrangian at zero
momentum.  This equivalence has been demonstrated for the fermionic
partition function, but is only conjectured for the partially quenched
case, which is related to the eigenvalues of the Dirac operator.  Here
we will directly compare estimates of the size of the taste breaking
in lattice simulations determined from the staggered chiral Lagrangian
(in the $p$-regime) with estimates obtained from comparing low
eigenvalues to the SRMT predictions (in the $\epsilon$-regime).
Our initial tests show good agreement with the predictions of SRMT
in support of the conjectured equivalence of the partially quenched
theories.

\section{Staggered Chiral Lagrangian}

The effective chiral Lagrangian for staggered fermions at order $a^2$
 is given by \cite{Lee:1999zxa,Aubin:2003mg}
\be
\label{cV}
\cL = \frac{F^2}{8} \tr{\partial_\mu U \partial_\mu U^\dagger} 
-\frac{1}{2} \Sigma_0 m \tr{U + U^\dagger}
+ a^2 \cV
\ee
where $\tr{X}$ stands for the trace of $X$, and
$F$ and $\Sigma_0$ are the low energy constants (LECs) related to the
pion decay constant (with the convention that the physical value for
 $F \approx 131$ MeV)
and the magnitude of the chiral condensate, respectively.
The taste breaking terms can be divided into two parts $\cV =
\cV_{1t} + \cV_{2t}$.  The first part contains the single-trace terms
(with $\xi_\mu = \gamma_\mu^*$)
\be
\label{sL}
-\cV_{1t} &=&
  C_1 \tr{ \xi_5 U \xi_5 U^\dagger }
+ C_3 \frac{1}{2} \sum_\mu \left[ \tr{ \xi_\mu U \xi_\mu U } + h.c. \right] \nl
&+& C_4 \frac{1}{2} \sum_\mu \left[ \tr{ \xi_{\mu 5} U \xi_{5 \mu} U } + h.c. \right]
+ C_6 \sum_{\mu<\nu} \tr{ \xi_{\mu\nu} U \xi_{\nu\mu} U^\dagger }
\ee
and the second part has the two-trace terms
\be
\label{c25tr2}
-\cV_{2t} &=& 
    C_{2V} \frac{1}{4} \sum_\mu \left[ \tr{ \xi_\mu U } \tr{ \xi_\mu U } + h.c. \right]
+ C_{2A} \frac{1}{4} \sum_\mu \left[ \tr{ \xi_{\mu5} U } \tr{ \xi_{5\mu} U } + h.c. \right] \nl
&+& C_{5V} \frac{1}{2} \sum_\mu \left[ \tr{ \xi_\mu U } \tr{ \xi_\mu U^\dagger } \right]
+ C_{5A} \frac{1}{2} \sum_\mu \left[ \tr{ \xi_{\mu5} U } \tr{ \xi_{5\mu} U^\dagger } \right]
.
\ee

\section{Staggered Chiral Random Matrix Theory}

The (fermionic) staggered chiral random matrix theory partition
function can be written as \cite{srmt}
\be
\label{SRMT}
Z_{SRMT} =
\int dW p_0(W) p_T(T) \prod_{f=1}^{N_f} \det(D+m_f)
\ee
with
\be
D = \left( \begin{array}{cc}
0 & i W \\
i W^\dagger & 0
\end{array} \right) 
\otimes \1_4 + a T  ~.
\ee
where $W$ is a $(N+\nu) \times N$ complex matrix with $\nu$ the absolute
 value of the topological charge and $T$ incorporates the taste breaking terms.
The matrix potential for $W$ is conveniently a Gaussian,
\be
p_0(W) = \exp(-\alpha N \tr{W^\dagger W})
\ee
with $\sqrt{\alpha} = \Sigma_0 V/ 2N$ ($V$ is the four volume).

The taste breaking contribution to the SRMT ($T$) has eight terms that correspond directly
with the eight taste breaking terms of the chiral Lagrangian.
Its complete form was given in \cite{srmt}.
As an example, the $C_4$ term corresponds to
\be
\label{t4}
T_4 = \sum_{\mu} \left( \begin{array}{cc} A_\mu & 0 \\ 0 & B_\mu \\ \end{array} \right)
\otimes \xi_{\mu 5}
\ee
where $A_\mu$ and $B_\mu$ are Hermitian matrices of size $(N+\nu) \times (N+\nu)$ and 
$N\times N$, respectively.
For convenience we can choose a Gaussian weight function for these matrices,
\be
p_{T_4} = \exp\left(-\frac{\alpha N^2}{2 V C_4} \sum_\mu \tr{A_\mu^2} + \tr{B_\mu^2} \right) ~.
\ee
One can then show that the (fermionic) RMT with this extra matrix term is equivalent
to the zero-momentum staggered chiral Lagrangian with just a $C_4$ taste breaking term.

Note that while the correction to the RMT enters at order $a$, this still reproduces
a term of order $a^2$ in the chiral Lagrangian.  This is due to the taste breaking terms
in the SRMT being traceless.  As demonstrated in \cite{srmt}, when expanding the
determinant of the SRMT Dirac matrix, the $O(a)$ term vanishes for this reason, and results
in a partition function that has a leading corrections at $O(a^2)$ even
 though the SRMT Dirac operator has terms of $O(a)$.

The two-trace terms can be incorporated in the SRMT in a couple of ways.
One is to add terms such as
\be
 \left( \begin{array}{cc} b_\mu \1_{N+\nu} & 0 \\
 0 & \pm b_\mu \1_{N} \\
 \end{array} \right) \otimes \xi_{\mu 5}
\ee
with a Gaussian weight for the scalar $b_\mu$.
This will give a contribution to the $C_{2A}$ and $C_{5A}$ terms.
While this term will reproduce the correct term in the chiral Lagrangian,
it has no analogue in the lattice Dirac matrix.  This term is of the
form of a fluctuating taste-dependent mass, which is not present on
the lattice.

An alternative form for the two-trace terms is to simply modify the potentials
for the matrix terms corresponding to the one-trace terms.  By replacing the simple
Gaussian weight with one that also includes $\tr{A_\mu}^2$, $\tr{B_\mu}^2$ and
$\tr{A_\mu}\tr{B_\mu}$ in the exponential, the coefficients can be tuned to give
the correct two-trace terms in the chiral Lagrangian \cite{srmt}.

\section{Generalized Staggered Random Matrix Theory}

If we look at the final form of the SRMT Dirac matrix, we see that it
is the most general matrix that is consistent with the symmetries of
the staggered Dirac operator; it is anti-Hermitian and anticommutes
with the staggered chiral symmetry matrix $\gamma_5 \otimes \xi_5$.
As mentioned in the previous section, the single-trace terms from the
chiral Lagrangian are reproduced by considering independent Gaussian
weights for the remaining matrix elements.  Meanwhile the two-trace
terms can be reproduced by adding two-trace terms to the RMT weights.

In this manner one could consider generalizing the SRMT to include
more terms in the weight function in an attempt to reproduce higher
order terms in the chiral Lagrangian (at zero momentum).  One can then
draw an analogy between the formulation of a RMT and of an effective
Lagrangian.  For the Lagrangian one includes all terms, up to some
order, that are consistent with the symmetries.  Likewise for the RMT
one can consider a matrix containing all elements consistent with the
symmetries of the Dirac operator, with a generalized weight function
for these matrix elements up to some level of complexity.  Of course the
mapping from the RMT potential to the chiral Lagrangian at higher
order may not be as simple as in the SRMT presented here, but one
could speculate that a generalized RMT could reproduce all higher
order terms of the chiral Lagrangian.  Again this equivalence
only holds for the zero momentum Lagrangian, but it is also possible to
reduce the full Lagrangian to an effective zero momentum Lagrangian
(for a recent example see \cite{Lehner:2010mv}),
which then might be representable as a RMT.

\begin{figure}[t]
  \begin{center}
    \includegraphics[width=0.8\textwidth]{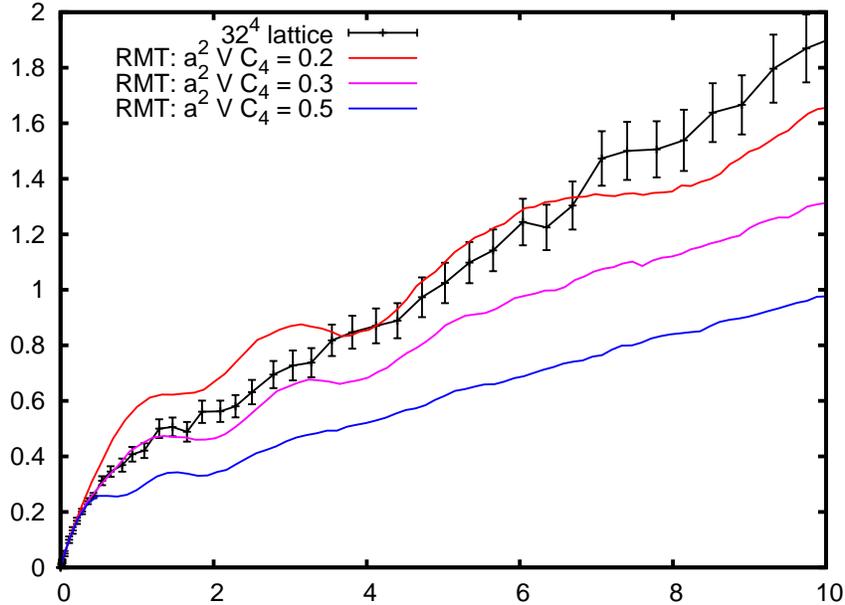}
    \caption{\label{nv}
      Number variance of the lowest eigenvalues of the 
      $32^4$ lattice ensemble and the SRMT.
    }
  \end{center}
\end{figure}

\section{Dominant form of taste breaking}

Most of the taste breaking coefficients can be measured in lattice
simulations by comparing to staggered chiral perturbation theory
(\scpt) results.  The four single-trace parameters are uniquely
determined from the splittings of the pion masses at leading order
\cite{Aubin:2003mg}.
Typically the $C_4$ term is found to be the dominant contribution to
the pion spectrum \cite{Lee:1999zxa}.  The two-trace terms enter in
\scpt{} formulas in the combinations
$C_{V,A}^{\pm} = (C_{2\{V,A\}} \pm C_{5\{V,A\}})/2$.
They don't contribute to the pion splittings at leading order,
but the $C_A^-$ and $C_V^-$ terms do appear in one-loop expressions,
while the $C_A^+$ and $C_V^+$ terms do not \cite{Aubin:2003mg}.
In lattice simulations $C_A^-$ has been found to be larger than $C_V^-$
\cite{Aubin:2004fs}.

Both of the dominant terms, $C_4$ and $C_A^-$, come from same term in
the SRMT with the more general form for the weight function.  This
supports the idea of constructing the SRMT from a single matrix with
the correct symmetries and with a generalized weight function.  Based
on the lattice measurements, the leading contribution to taste
breaking in the staggered Dirac matrix has an axial-vector taste
structure.  If one imagined rotating the staggered Dirac matrix into a
taste basis and then expanding in powers of $a$, the dominant
correction at order $a$ would then have the same form as in
(\ref{t4}).  Of course the exact potential for the lattice Dirac
matrix would be much more complicated than in the SRMT, but the
leading effects at low energy can be captured by the SRMT weight
function considered here.  Among the terms in the SRMT potential
corresponding to the axial-vector matrix, we have no reason to favor
one over the other, so we would naively expect them to be of
similar magnitude.  We would then expect the corresponding terms
in the chiral Lagrangian to be of similar order, and also
dominant over the corresponding terms with other taste symmetries,
which is consistent with lattice measurements.

\section{Extracting LECs from RMT}

The RMT predictions for the eigenvalue correlations can be used to
extract LECs from lattice simulations.
For example $\Sigma_0$ can be obtained by fitting to the eigenvalue
density.
Additionally $F$ can be obtained from the correlations of eigenvalues
with an imaginary chemical potential \cite{rmtf}.

If the taste breaking is small enough then these methods can apply
directly to staggered eigenvalues by replacing each quartet of
eigenvalues with its average \cite{latev}.  In this case one could also try
to extract the taste breaking parameters from the splittings of the
eigenvalues within the quartet.  In practice, it would likely be too
difficult to extract all the parameters, but if one assumes that the
$C_4$ term is dominant, then it should be possible to estimate it from
comparison to the SRMT.

However, if the taste breaking is large, then the higher order taste
breaking terms can become important.  In this case the effective
chiral Lagrangian reduces to a single flavor for the remaining
staggered chiral symmetry with a new set of LECs that are in principle
unrelated to the original ones \cite{srmt}.  Thus extracting the LECs
from low eigenvalues when there aren't clear quartets present may not
yield the continuum LECs in chiral Lagrangian.

\begin{figure}
  \vspace{-7mm}
  \begin{center}
    \begin{minipage}[t]{.49\textwidth}
      \includegraphics[clip,width=\textwidth]{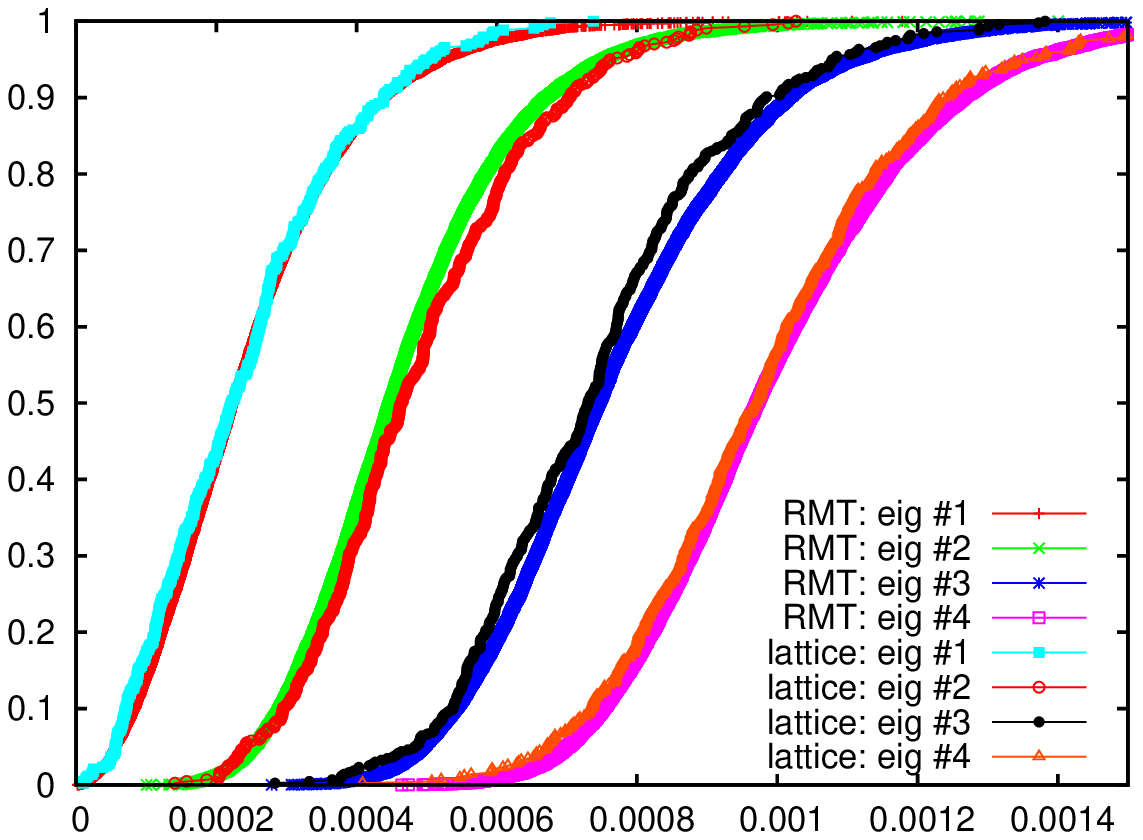}
    \end{minipage}
    \begin{minipage}[t]{.49\textwidth}
      \includegraphics[clip,width=\textwidth]{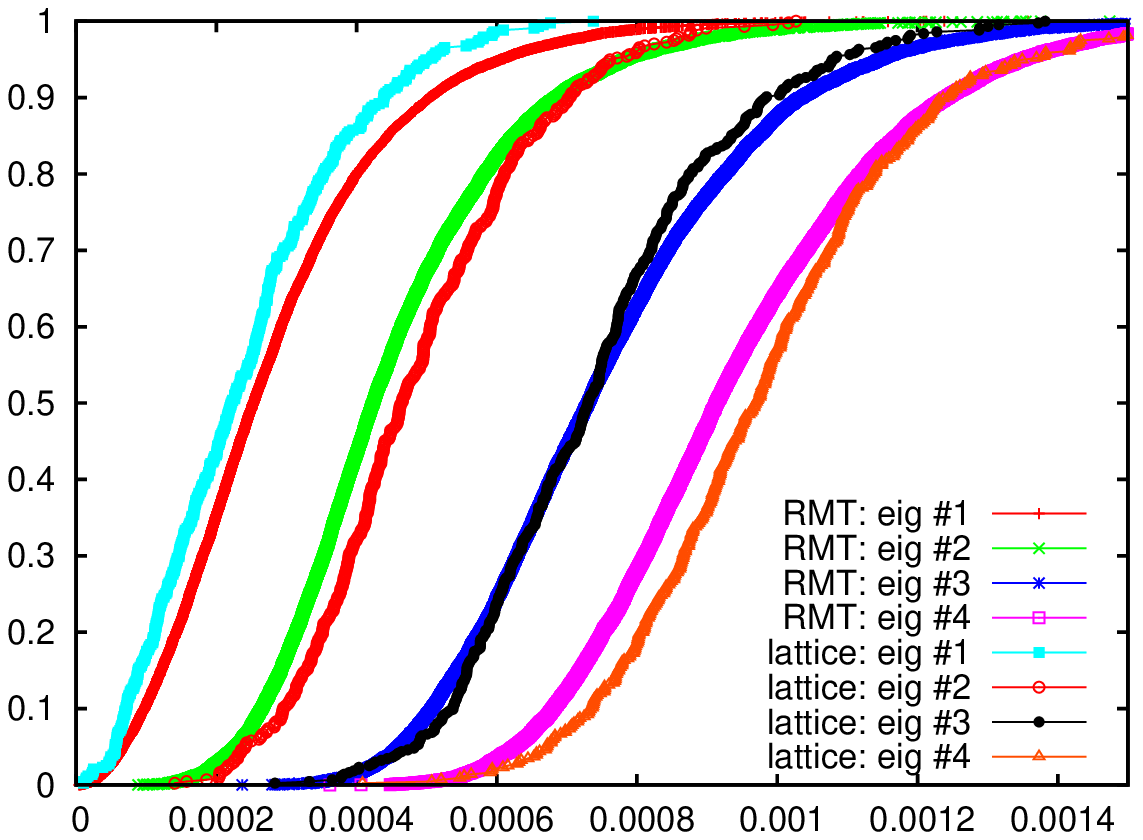}
    \end{minipage}
    \caption{\label{id}
      Individual integrated densities of the lowest four eigenvalues
      from the $32^4$ lattice ensemble and the SRMT.  The dominant taste breaking
      parameter in the SRMT is set to $a^2VC_4 = 0.3$ (left) and $0.2$ (right).
    }
  \end{center}
  \vspace{-7mm}
\end{figure}

\section{Comparison to lattice simulations}
\vspace{-2mm}

As an initial test of the SRMT, we will compare the leading order taste breaking
term obtained from fitting the Dirac operator eigenvalues to the SRMT with that
obtained from fitting the pion spectrum to the staggered chiral Lagrangian.
Since it is difficult to get a single lattice ensemble that we could use for
both measurements we will use two ensembles with all parameters identical except
for the volume.

For the pion masses we use an ensemble from the MILC collaboration
2+1+1 flavor HISQ runs \cite{Bazavov:2010pi}.  The ensemble has a
volume of $32^3\times 96$ with a lattice spacing of $a\approx 0.09$ fm and
with a light quark mass $m_l = m_s/5$.  From the pion mass splittings
we can get the single-trace taste breaking terms in the chiral
Lagrangian.  Using a value of $F=131$ MeV, this gives
\be
a^2 V C_1 =  0.03(8),~~
a^2 V C_3 = -0.03(4),~~
a^2 V C_4 =  0.84(4),~~
a^2 V C_6 =  0.03(3)  ~.
\ee
We can see here that $C_4$ is clearly dominant, as expected, and that
the other coefficients are all consistent with zero.

We generated a new ensemble of 430 lattices of size $32^4$ with all
other parameters the same as the previous ensemble.  From the volume
scaling we expect to find
\be
\label{c4}
a^2 V C_4 = 0.28(1)
\ee
on this new ensemble.
We then compare lattice eigenvalues with numerical simulations of the SRMT
with only the $C_4$ taste breaking term.  For this comparison we choose to
use the number variance statistic.  This shows the fluctuations
(variance) in the number of eigenvalues in an interval starting at
zero versus the average number in the interval.  We evaluated it
numerically from simulations of the SRMT with $N=400$.

In Figure \ref{nv} we plot the number variance of the lattice
eigenvalues against the SRMT at different values of $a^2 V C_4$.  We
can see that the best agreement for small intervals (up to around
2 to 4 eigenvalues) is at $a^2 V C_4 \approx 0.3$.
This is in good agreement with our prediction (\ref{c4})
obtained from the pion mass splittings.
As the length of the interval grows,
the lattice results start to move away from the SRMT result.  This is
likely due to higher momentum modes entering on the lattice, that aren't
captured in the RMT.  This happens at the QCD equivalent of the Thouless energy \cite{te},
which for this ensemble, in units of the average eigenvalue spacing,
is $F^2 \sqrt{V} \approx 3.5$.
This is consistent with our observations from the number variance.

As a further check that the SRMT describes the low eigenvalues of the
staggered Dirac operator, we look at the individual integrated
densities of the lowest four eigenvalues.
In Figure \ref{id} we see the integrated density from the $32^4$
lattice along with that from the numerical simulations of the SRMT at
two different values of the taste breaking parameter
$a^2VC_4 = 0.2,0.3$.
We can see that the value of $0.3$ fits the lattice data much better
than at $0.2$, again confirming that the predictions of the SRMT
fit the lattice data well and give estimates of the taste breaking
parameters that are consistent with the staggered chiral Lagrangian.

\section{Summary}

We have shown a chiral RMT that incorporates all leading order taste
 breaking terms from staggered chiral Lagrangian.
The SRMT can be constructed by considering a RMT
 with same symmetries as the staggered Dirac operator and a generalized weight function.
Initial tests show that the predictions of the SRMT are in good agreement with
 lattice simulations within the range of validity of the SRMT.
Additionally, using the predictions of the SRMT,
 the dominant taste breaking parameter can be extracted from
 the low eigenvalues of the staggered Dirac operator and gives consistent
 results with that obtained from the pion mass splittings.

\begin{acknowledgments}
  We thank Doug Toussaint for providing the pion splitting data for
  the MILC ensemble.  This research used resources of the Argonne
  Leadership Computing Facility at Argonne National Laboratory, and
  was supported by the U.S. DOE under contract DE-AC02-06CH11357.
\end{acknowledgments}

\end{document}